\documentclass{article}

\usepackage{PRIMEarxiv}

\usepackage[utf8]{inputenc} 
\usepackage[T1]{fontenc}    
\usepackage{hyperref}       
\usepackage{url}            
\usepackage{booktabs}       
\usepackage{amsfonts}       
\usepackage{nicefrac}       
\usepackage{microtype}      
\usepackage{lipsum}
\usepackage{fancyhdr}       
\usepackage{graphicx}       
\graphicspath{{media/}}     
\usepackage{booktabs}
\usepackage{float}
\usepackage{color}

\title{The Ethics of AI-Generated Maps: A Study of DALL·E 2 and Implications for Cartography
}

\author{
  Yuhao Kang \\
  GeoDS Lab, Department of Geography, University of Wisconsin-Madison \\
  \texttt{yuhao.kang@wisc.edu} \\
   \And
  Qianheng Zhang \\
  HGIS Lab, Department of Geography, University of Washington \\
  \texttt{qianhz2@uw.edu} \\
  \AND
  Robert Roth \\
  University of Wisconsin Cartography Lab, Department of Geography, University of Wisconsin-Madison \\
  \texttt{reroth@wisc.edu} \\
}

\begin{document}
\maketitle

\begin{abstract}
The rapid advancement of Artificial Intelligence (AI), such as the emergence of large language models ChatGPT and DALL·E 2, has brought both opportunities for improving productivity and raised ethical concerns.
This paper investigates the ethics of using generative AI in cartography, with a particular focus on the generation of maps using DALL·E 2. 
We ask if we can use GeoAI techniques to identify synthetic maps generated by the DALL·E 2.
To accomplish this, we first created an open-source dataset that includes synthetic (AI-generated) and real-world (human-designed) maps at multiple scales with a variety settings.
We subsequently examined four potential ethical concerns that may arise from the characteristics of DALL·E 2 generated maps, namely inaccuracies, misleading information, unanticipated features, and irreproducibility.
We then developed a deep learning-based AI-generated map detector to identify those AI-generated maps.
Our research emphasizes the importance of ethical considerations in the development and use of AI techniques in cartography, contributing to the growing body of work on trustworthy maps.
We aim to raise public awareness of the potential risks associated with AI-generated maps or potential future maps generated by Artificial General Intelligence (AGI) and support the development of ethical guidelines that govern their future use.
We have made our dataset publicly available on GitHub: \href{https://github.com/GISense/DALL-E2-Cartography-Ethics}{https://github.com/GISense/DALL-E2-Cartography-Ethics}.
\end{abstract}

\keywords{DALL·E 2 \and Ethics \and Cartography \and Map}

\section{Introduction}
Cartographers long have recognized the significance of developing ethical and trustworthy maps, i.e., maps that truthfully depict geographic information while minimizing the introduction of misinformation or bias \cite{monmonier2018lie,griffin2020trustworthy}.
With the rapid advancements in Artificial Intelligence (AI), the use of AI in map-making has brought both opportunities and concerns \cite{mai2023opportunities, uhl2021combining, kang2022review, chiang2020training}.
On the one hand, (Geo)AI techniques can facilitate map creation processes and even have demonstrated the potential to support human creativity in cartographic design. 
For instance, cartographers have employed (Geo)AI to support cartographic design decisions on the artistic aspects of maps such as map style transfer \cite{kang2019transferring,christophe2022neural}, map retrieval \cite{robinson2022visualizing}, map generalization \cite{feng2019learning,courtial2023deriving, shen2022raster, yan2019graph}, and map design critique \cite{chen2021learning}.
On the other hand, despite its promise, cartographers have expressed ethical concerns about the uncertainty and opacity (i.e., machine learning and deep learning models often are considered as ``black-boxes'') of AI for generating maps \cite{zhao2021deep,kang2022review}.
As \cite[p.~9]{griffin2020trustworthy} asks: ``How much should we trust a machine-generated map?''

Starting in November 2022, generative AI models such as ChatGPT and DALL·E 2 have attracted significant public attention \cite{openai2023gpt4, van2023chatgpt}.
These large language models (LLM) have demonstrated impressive capabilities in tasks such as language generation and image synthesis.
Several researchers have suggested that existing generative AI models like ChatGPT could potentially serve as an initial prototype or precursor to the development of Artificial General Intelligence (AGI), which refers to an AI system with the capacity to perform intellectual tasks equivalent to humans \cite{bubeck2023sparks}.
Yet, they have also fueled debates surrounding the ethical concerns related to the development of generative AI and AGI \cite{mclean2021risks,zhuo2023exploring, liebrenz2023generating}.
While these models hold the potential to revolutionize several industries, their advancement also raises critical questions about the future of labor \cite{zarifhonarvar2023economics}, trust \cite{tlili2023if}, and the consequences of unbridled technological development \cite{van2023chatgpt, mhlanga2023open, king2023conversation}. 
Therefore, there is an urgent need for careful consideration and ethical evaluation of AI technologies in myriad domains to ensure their responsible and beneficial use.

As cartographers and geographers, we have a particular interest in investigating the ethical implications of maps created by these advanced generative models.
The emergence of powerful tools such as DALL·E 2 has made it increasingly accessible to generate high-quality map images by providing specific prompts.
However, this also has introduced new challenges related to the accuracy and trustworthiness of these synthetic maps generated by AI. 
While these maps may look realistic, they also may contain inaccuracies or be influenced by biases embedded in the AI models, resulting in the proliferation of potentially meaningless, and, at worse, harmful maps \cite{griffincall}.
To address these issues, it is necessary to build solutions for detecting and mitigating the risks associated with using such maps, as suggested by \cite{zhao2021deep}. 
Hence, it is crucial to offer timely detection of ``fake'' maps to assess the trustworthiness of web maps and minimize the potential negative impacts associated with their use.
The significance of this task is even more critical due to the enormous amount of information that is available on online news and social media platforms every day.

To this end, we aim to investigate the use of AI in generating maps and the associated ethical implications of AI-generated maps.
We ask the following two fundamental questions: (1) What potential ethical concerns arise from the characteristics of maps generated by DALL·E 2? and (2) How can AI-generated maps be identified as an initial step towards minimizing their potential ethical concerns?
To accomplish this, we first create a dataset that contains synthetic maps generated by DALL·E 2 with diverse prompts at multiple spatial scales (hereafter referred to as \textit{AI-generated maps}).
We also collect real-world maps using search engines (hereafter referred to as \textit{human-designed maps}).
After offering the implications of using AI in map generation for cartography, we train a deep learning-based model capable of identifying AI-generated maps.
To the best of our knowledge, this is among the first studies to apply ChatGPT-like models (e.g., DALL·E 2) in cartography.
Our research contributes to the growing body of work on trustworthy maps and the ethics of cartography, highlighting the importance of ethical considerations in the development and use of AI techniques for cartography.
By exploring these issues, we aim to raise public awareness about the potential risks associated with AI-generated maps and foster the development of future ethical guidelines for their use, both within and outside the cartography, geography, and GIScience community.

\begin{figure}[H]
    \centering
    \includegraphics[width=\textwidth]{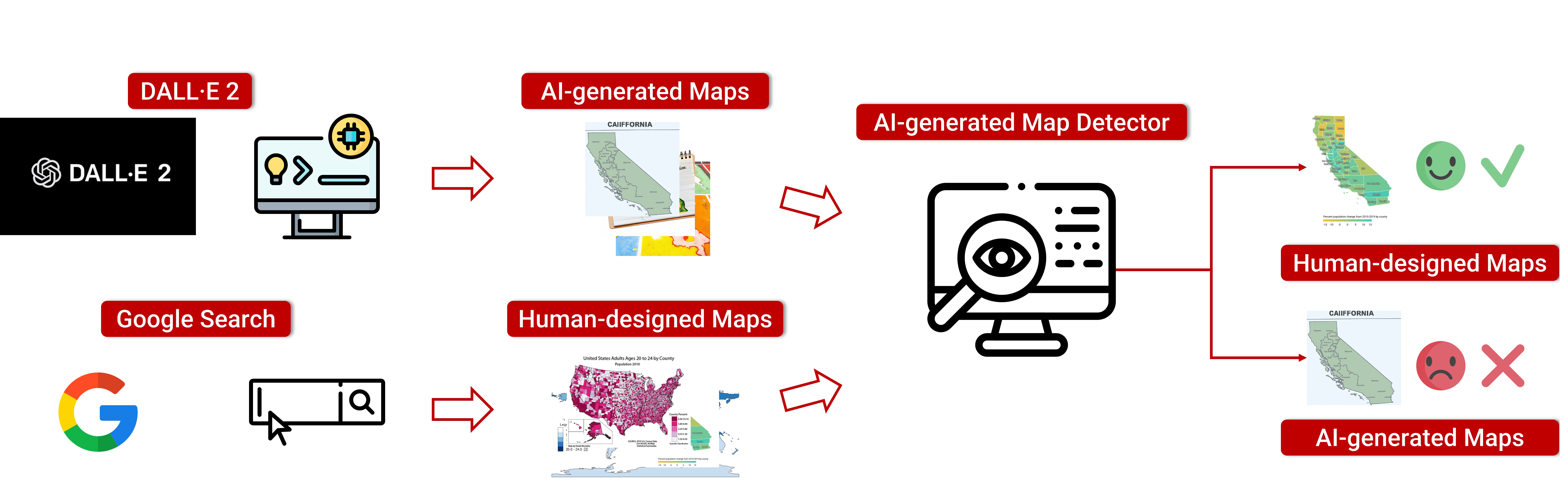}
    \caption{\textit{The computational workflow for identifying AI-generated maps.}}
    \label{fig:framework}
\end{figure}

\section{Data and Methodology}
\label{sec:headings}
Figure \ref{fig:framework} shows the computational framework for this study.
First, we generated a dataset that includes both AI-generated and human-designed maps, which can further be used to train and evaluate deep learning models for identifying AI-generated maps.
The AI-generated maps were created using DALL·E 2, and human-designed maps were obtained online via a search engine.
Second, we developed a deep learning-based AI-generated map detector.
It performs a binary classification to judge whether a map is a human-designed map or an AI-generated map.
Such a proposed computational framework may help cartographers to analyze AI-generated maps and contribute to the trustworthiness of maps.

\subsection{Dataset Construction}
\subsubsection{AI-generated Maps}
\begin{figure}[H]
    \centering
    \includegraphics[width=0.60\textwidth]{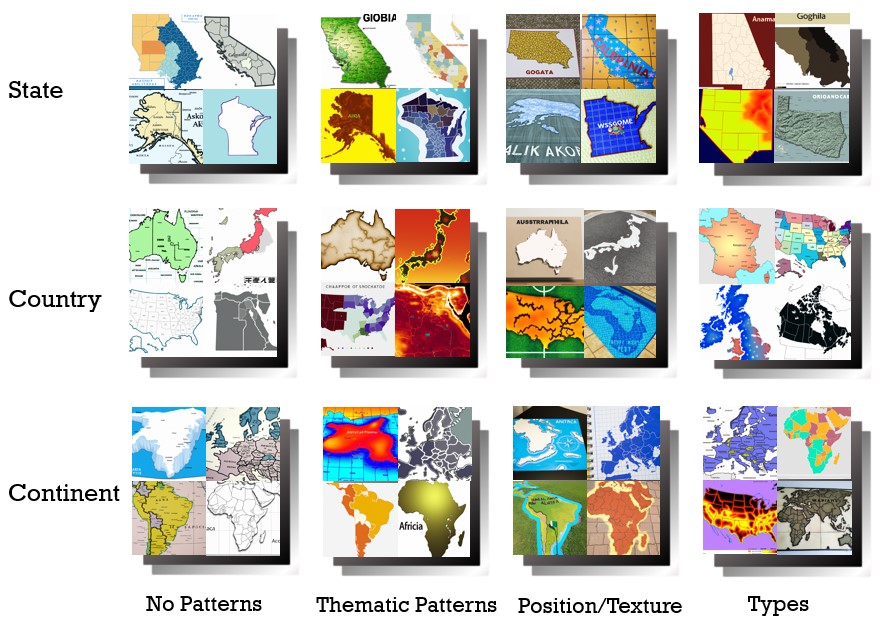}
    \caption{\textit{Example maps of AI-generated maps using DALL·E 2. Region levels include States, countries, and continents. Descriptive prompts include no pattern(Maps without specific thematic or positional prompts), thematic patterns, positional patterns, and map types(reference, physical, choropleth, etc..).}}
    \label{fig:examples}
\end{figure}

We first create an AI-generated map dataset using the DALL·E 2\footnote{\href{https://openai.com/product/dall-e-2}{https://openai.com/product/dall-e-2}}.
DALL·E 2 is an advanced generative model developed by OpenAI that relies on prompts to generate images. 
Unlike ChatGPT, which excels at text-to-text generation, DALL·E 2 was trained on a dataset of text-image pairs, allowing it to generate realistic and diverse images from textual prompts. 
Figure \ref{fig:examples} displays several example maps that are generated by the DALL·E 2, showcasing their diverse styles, layouts, and spatial patterns.
Specifically, we generate the maps using the following prompt format:

\begin{center}
``\textit{A \{MapType\} of \{Region\} on \{Place\} with \{Description\}}''
\end{center}

This format allowed us to specify the map type, mapped region, the location of the map within the image, and additional descriptive information for the AI model to generate a corresponding map.
The \textit{MapType} parameter included six categories: choropleth map, general map, heat map, physical map, political map, and reference map.
To account for the different scales at which maps are displayed, we generated maps at three levels of \textit{Region}: continent, country, and state. 
At the continent level, we included all seven continents around the world;
at the country level, we selected 100 countries based on GDPs;
for state-level maps, we focused on the 50 states of the United States.
The first two parameters for all prompts, \textit{MapType} and \textit{Region}, were required, while \textit{Place} and \textit{Description} were optional. 
\textit{Place} refers to the location of the map within the image, which could be on a table, floor, desk, or football field, rather than the image being a map in its entirety; the full list of places is provided in the Appendix Table \ref{tab:appendix}. 
We also included a range of \textit{Description} options for generating choropleth and heat maps with different styles, such as bright colors, blue and green hues, and more. 
For instance, to generate a United States choropleth map that is in warm colors, the prompt could be: ``A choropleth map of United States with warm colors''.
Next, we randomly selected options for each parameter and generated a diverse set of maps covering various regions and themes. 
Such a comprehensive dataset was used for further analysis and training in the deep learning-based AI-generated map detector.
Table \ref{tab:num_maps} summarizes the number of maps generated by AI at different scales.
We have made the dataset openly available on GitHub at: \href{https://github.com/GISense/DALL-E2-Cartography-Ethics}{https://github.com/GISense/DALL-E2-Cartography-Ethics}.

\subsubsection{Human-designed Maps}
As a comparison, we developed a Python web scrapper to collect maps from the Google search engine at the same levels and administrative regions.
To do so, we entered a search query in the format ``\textit{\{Region\}} maps'' on April 4, 2023, such as ``United States maps'', where \textit{Region} refers to the same scales and administrative regions as in the AI-generated map dataset.
We then collected several top images (ranging from 50 to 100 depending on the scales) among all the returned images for each search query.
We adopted such a strategy used in prior studies to construct map datasets for collecting images at the country and continent levels \cite{evans2017livemaps,hu2022enriching}.
Regarding images at the state level, we directly utilized the dataset released from \cite{hu2022enriching}.
Table \ref{tab:num_maps} summarizes the number of map images that we collected from Google at different scales.

\begin{table}[H]
\centering
\caption{\textit{Number of maps}}
\label{tab:num_maps}
\begin{tabular}{@{}lllll@{}}
\toprule
                                   & State (50 states)                           & Country (100 countries)                         & Continent (7 continents)                     & Total                     \\ \midrule
AI-generated Dataset & 30 images * 50 =1500 & 30 images * 100 =3000 & 25 images * 7 =175 & 4675 \\
Human-designed Maps  & 50 images * 50 =2500                      & 50 images * 100 =5000                      & 100 images * 7 =700                     & 8200                      \\ \bottomrule
\end{tabular}
\end{table}

\subsection{Development of the AI-generated Map Detector}
Based on the datasets, we developed an AI-generated map detector that can identify AI-generated maps which may offer potential solutions for creating trustworthy maps.
Such an AI-generated map detector is developed based on a ResNet model \cite{he2016deep}. 
The ResNet model is a Deep Convolutional Neural Network (DCNN) that has been widely used in computer vision tasks due to its outstanding performance.
One of the unique features of ResNet is its ability to handle a DCNN with a depth of up to hundreds of layers. 
ResNet has proven to be highly effective in image classification tasks.
Given its mature and high accuracy in image classification, we utilized the ResNet model in our study to classify maps as either generated by AI or created by humans.
To train the model, we combined the two datasets, namely, AI-generated maps and human-designed maps, and input them into the ResNet model.
This can be viewed as a binary classification to determine whether an input map is generated by AI or by humans.
The dataset was split into training, validation, and test sets, with 70\%, 15\%, and 15\% of the data, respectively. 
We trained the ResNet-18 model on the training set and evaluate the performance on the validation dataset. 
It should be noted that, to be fed into the ResNet model, we prepared the dataset by resizing all maps to size 224x224. 
Then the ResNet model was trained with SGD optimizer and the cross-entropy loss function. 
Once the model converged, we used the output model to predict whether an input map was generated by AI or by humans.

\section{Results}
\subsection{Ethical Concerns of AI-generated maps}
\begin{figure}[h]
    \centering
    \includegraphics[width=\textwidth]{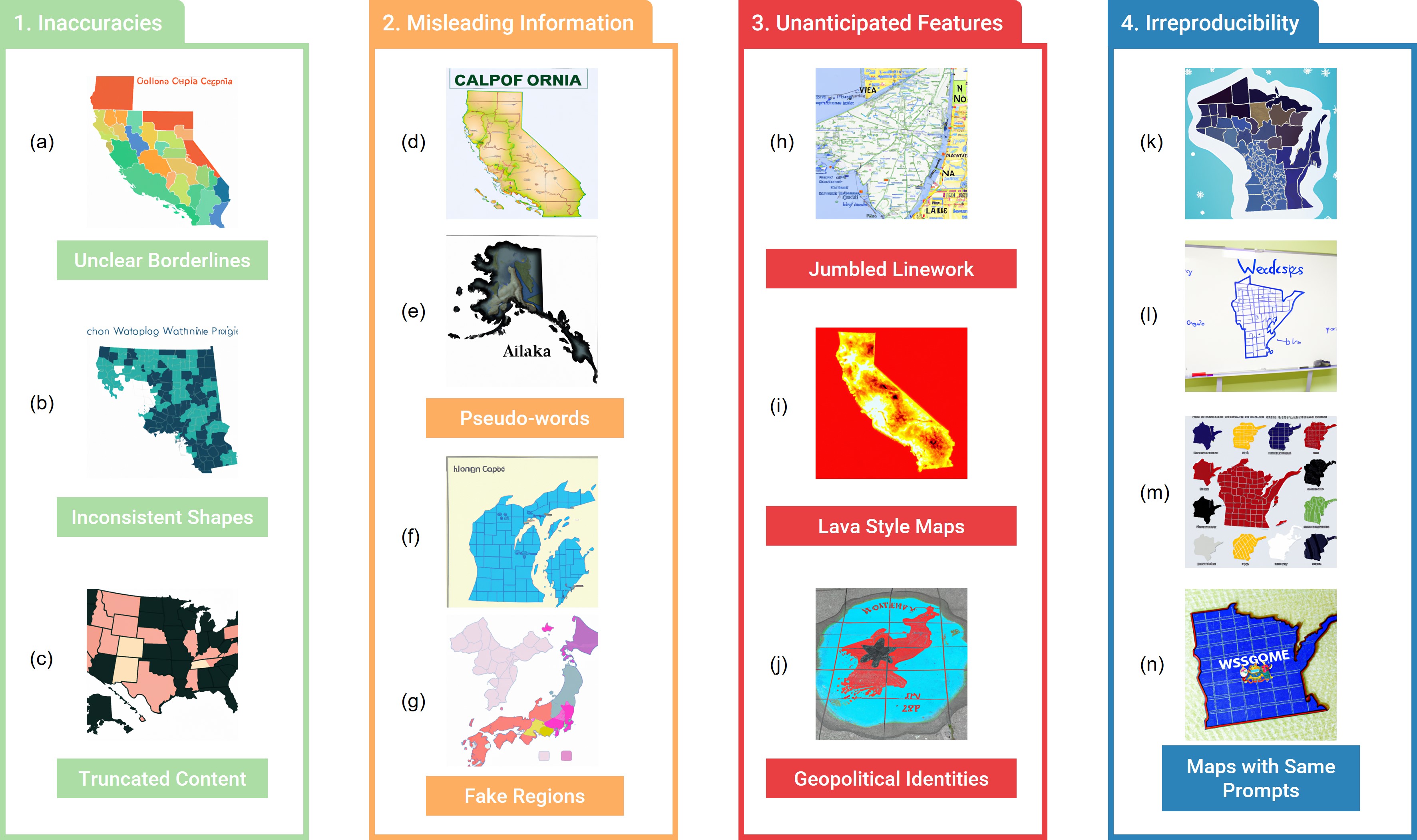}
    \caption{\textit{Example AI-generated maps: (1) inaccuracies, (2) misleading information, (3) unanticipated features, and (4) irreproducibility.}}
    \label{fig:example_dalle}
\end{figure}

Based on our qualitative observations of AI-generated maps, we summarize four potential ethical concerns of such maps: inaccuracies, misleading information, unanticipated features, and irreproducibility.
Figure \ref{fig:example_dalle} shows several examples of these map characteristics from the four aspects.

The inaccuracies observed in AI-generated maps are primarily manifested through three forms: unclear borderlines, inconsistent shapes of places, and truncated content.
Specifically, AI-generated maps may have unclear and distorted borderlines between different regions (e.g., states, and counties).
Figure \ref{fig:example_dalle}a portrays an example of a distorted map, shows California, United States, with distorted borders between counties.
The shapes of places also may be inconsistent, with the shape of a particular region varying from one map to another.
Figure \ref{fig:example_dalle}b depicts a map of the state of Washington in the United States, which exhibits significant distortions in its shape.
Interestingly, certain places may have a higher probability of shape variation across maps compared to others.
For instance, we anecdotally observed greater variation in Wisconsin's outer shapes than California's in AI-generated maps.
A possible explanation for this might be the training dataset size, with more maps available for California than for Wisconsin.
Furthermore, Figure \ref{fig:example_dalle}c shows truncated context due to the fixed square aspect ratio of DALL·E 2 output images.
For instance, most maps depicting the United States only showcase specific regions of the contiguous territory, while omitting certain boundaries (e.g., part of the West and East coastal areas, Alaska, and Hawaii).

AI-generated maps also can produce misleading information. 
For example, AI-generated maps in Figure \ref{fig:example_dalle}d-e include pseudo-words, symbols, or characters that resemble the names of countries or cities, potentially leading to confusion or misinterpretation. 
Moreover, AI-generated maps also can contain fake countries, states/provinces, or counties that do not actually exist in reality, creating a false impression of the current map. 
For instance, Figure \ref{fig:example_dalle}f presents a map of Michigan, United States, with a nonexistent region to the east, while Figure \ref{fig:example_dalle}g depicts a map of Japan with nonexistent territory to the west.
These features potentially can lead to the spread of misinformation, which may distort popular notions of reality and have unintended geopolitical consequences.
As such, researchers and practitioners in cartography, geography, and GIScience must consider the impact of these features and work to minimize the potential negative effects through careful evaluation and scrutiny of potentially AI-generated web maps.

In addition to the inaccuracies and misleading information described earlier, AI-generated maps may create unexpected or unanticipated features.
For instance, AI-generated maps are unaware of the underlying geographic processes that lead to repeated patterns in the landscape --- particularly for the built environment in our study --- resulting in distorted polygons and jumbled linework.
For instance, Figure \ref{fig:example_dalle}h displays jumbled road networks in the state of New York, United States.
Further, AI-generated maps may confuse established cartographic concepts and terminology.
Figure \ref{fig:example_dalle}i depicts a result for ``heat map'', with the output map styled in glowing tones similar to lava flows, suggesting that the model misclassified the meaning of the prompt.
In addition, AI models may generate specific themes of maps that reflect certain geopolitical identities, even if not input in keyword prompts.
For instance, Figure \ref{fig:example_dalle}j is a map of North Korea generated based on the prompt ``A physical map of North Korea on the pavement'' without explicitly mentioning its geopolitical identities.
However, the map is covered with the flags of North Korea, reinforcing its national identity.
Further work is needed to evaluate the degree to which AI-generated maps may stoke nationalism and thus reinforce xenophobic or otherwise biased geopolitical discourse.

Finally, AI-generated maps cannot be reproduced with the same keyword set. 
Due to the randomness inherent in the generation process of DALL·E 2, it is impossible to generate two maps that have the exact same content, map shapes, map styles, or overall layouts. 
For instance, Figure \ref{fig:example_dalle}k-n depicts Wisconsin maps generated using the same prompts ``A choropleth map of Wisconsin'', yet they exhibit diverse outer shapes, interior content, and visual styles.
Without greater reproducibility, cartographic research on GeoAI cannot be validated or replicated, and therefore poise ethical questions about the effectiveness of a convention, science-based peer-review system, and what ``counts'' as knowledge and scholarship more broadly \cite{wilson2021five}.
While it should be noted that, from a technical perspective, the model may reproduce the same outputs if we have the same hyperparameters (e.g., random seed, steps, prompts, weights), the model is not currently open-sourced and therefore is not reproducible at this time.

\subsection{Performance of the AI-generated Map Detector}

We evaluated the performance of our AI-generated map detector on the test set.
We computed the confusion matrix to measure the performance of the system, as illustrated in Table \ref{tab:confusion_matrix}.
Based on the results, we computed four commonly used metrics in machine learning, namely, accuracy, precision, recall, and F1 score, to measure the performance.
The system achieved an accuracy of 0.908, precision of 0.87, recall of 0.878, as well as an F1 score of 0.874 on the testing dataset.
These metrics suggest that the system is robust and effective in distinguishing between AI-generated maps and human-designed maps.

\begin{table}[h]
\centering
\renewcommand{\arraystretch}{1.3}
\caption{\textit{Confusion matrix for the performance of the AI-generated map detector on the testing dataset}}
\label{tab:confusion_matrix}
\begin{tabular}{ccc}
\hline
\textbf{} & \textbf{Actual AI-generated Maps} & \textbf{Actual Human-designed Maps} \\
\textbf{Predicted AI-generated Maps} & 616 & 92 \\
\textbf{Predicted Human-designed Maps} & 86 & 1135 \\
\hline
\end{tabular}
\end{table}

\section{Discussions and Conclusions}
While generative AI such as DALL·E 2 and ChatGPT have the potential to assist the cartographic design process, they also raise significant ethical concerns. 
In this paper, we presented an AI-generated map dataset using DALL·E 2 and investigated the potential ethical issues associated with AI-generated maps based on their characteristics.
The results suggest that even though AI-generated maps may offer potential benefits, they still may provide inaccurate or misleading information, possess unanticipated characteristics, and lack reproducibility.
Therefore, it is crucial that cartographers, geographers, and GIScientists must carefully consider these concerns when developing and using AI-generated maps to minimize the potential negative effects and ensure ethical and responsible usage.

We also developed an AI-generated map detector with deep learning that can identify whether a map is generated by AI or by humans.
Such an AI-generated map detector can be useful in various applications, such as identifying potential cases of AI-generated maps being used to spread misinformation on online social media platforms.
It may also significantly impact cartographic education, as students may use AI to generate maps for their assignments similar to how ChatGPT is increasingly employed for writing assignments \cite{rudolph2023chatgpt}.
Lecturers can apply such an AI-generated map detector to distinguish between maps generated by humans and machines.
Also, inaccurate or misleading maps, whether intentionally or unintentionally created, can have significant negative impacts, particularly in sensitive political or cultural contexts. 
The consequences of such maps can be severe, leading to fake impressions and disadvantages for already underprivileged groups.
Such an AI-generated map detector can help prevent the spread of misinformation and reduce the potential harm caused by AI-generated maps.

In addition, another potential ethical concern is related to commercialization and copyright.
Following the discussions in \cite{crampton1995ethics, kang2022review}, cartographers arguably should receive compensation if their maps are used as training data for generative AI models, but questions remain regarding who should be responsible for providing it. 
Moreover, if an AI-generated map is utilized for commercial purposes, the issue of compensation arises for the parties involved in its development, including the cartographers who provided training data and the company that developed the AI model. 
It is also worth noting that the training data used for generative AI models may come from publicly available sources or open data, which raises questions about who owns the copyright and how it should be managed.
These questions highlight the need for clear ethical guidelines and regulations related to the commercialization and copyright issues of the use of AI-generated maps.
Currently, DALL·E 2 users retain ownership of their generated images \footnote{\href{https://help.openai.com/en/articles/6425277-can-i-sell-images-i-create-with-dall-e}{https://help.openai.com/en/articles/6425277-can-i-sell-images-i-create-with-dall-e}}.
As such, the map dataset we provided in this paper is owned by the authors.


We acknowledge several limitations that worth examine in the future.
First, the dataset utilized in this paper was limited in terms of geographic coverage and diversity.
Collecting more maps with diverse characteristics (e.g., locations, styles) is necessary to enhance the generalizability of our findings. 
Second, this paper has only investigated the maps generated by DALL·E 2.
However, there have been numerous other models available, including but not limited to MidJourney\footnote{\href{https://www.midjourney.com/}{https://www.midjourney.com/}} and Stable Diffusion \cite{rombach2022high}, that can produce maps.
Future studies could incorporate these additional AI-generated data sources to expand the dataset and compare the maps generated by different models.

This manuscript is one of the first several works utilizing generative AI in cartography, we propose several promising directions for future research.
As suggested in the moonshot question in \cite{kang2022review}, ``\textit{Can we develop an artificial cartographer assistant so that cartographers are no longer focused on the usage of cartographic tools and technical details but more on artistic map creation?}''
the current development of AI may enable us to achieve this goal in cartography sooner than anticipated.

One potential future direction for AI in cartography refers to the generation of more accurate and visually appealing maps through the rapid evolution of (Geo)AI technology.
However, current generative AI technology may not have sufficient attention and focus on cartography, leading to limitations in handling cartographic tasks in comparison with other domains.
To address this issue, it is critical for cartographers to work alongside AI developers, and have the cartographer-in-the-loop in the development of AI, to ensure that the future AI and AGI could produce higher quality maps while minimizing potential ethical concerns and negative social implications.

Also, AI could be used to facilitate collaborative mapping efforts \cite{chambers2006participatory}. 
By integrating multiple data sources and input from various stakeholders, generative AI may help non-experts, even with limited text input, create maps, making maps more accessible to all people.
However, the potential ethical issues (e.g., bias, trustworthiness) should be monitored or reduced.
One possible solution refers to participatory mapping that allows people to contribute their local knowledge and expertise to the mapping process.
Incorporating participatory mapping into AI-generated maps may improve the accuracy of maps, which may also promote community engagement and collaboration.

\section*{Acknowledgments}
The authors would like to express their sincere gratitude for the technical support, valuable suggestions, and insightful discussions received from Dr. Song Gao at the GeoDS Lab, University of Wisconsin-Madison, Dr. Bo Zhao, and Yifan Sun at the HGIS Lab, University of Washington, Dr. Gengchen Mai at the University of Georgia.

\bibliographystyle{plainurl}  
\bibliography{references}

\begin{thebibliography}{10}

\bibitem{bubeck2023sparks}
S{\'e}bastien Bubeck, Varun Chandrasekaran, Ronen Eldan, Johannes Gehrke, Eric
  Horvitz, Ece Kamar, Peter Lee, Yin~Tat Lee, Yuanzhi Li, Scott Lundberg,
  et~al.
\newblock Sparks of artificial general intelligence: Early experiments with
  gpt-4.
\newblock {\em arXiv preprint arXiv:2303.12712}, 2023.

\bibitem{chambers2006participatory}
Robert Chambers.
\newblock Participatory mapping and geographic information systems: whose map?
  who is empowered and who disempowered? who gains and who loses?
\newblock {\em The Electronic Journal of Information Systems in Developing
  Countries}, 25(1):1--11, 2006.

\bibitem{chen2021learning}
Taisheng Chen, Menglin Chen, A-Xing Zhu, and Weixing Jiang.
\newblock A learning-based approach to automatically evaluate the quality of
  sequential color schemes for maps.
\newblock {\em Cartography and Geographic Information Science}, 48(5):377--392,
  2021.

\bibitem{chiang2020training}
Yao-Yi Chiang, Weiwei Duan, Stefan Leyk, Johannes~H Uhl, Craig~A Knoblock,
  Yao-Yi Chiang, Weiwei Duan, Stefan Leyk, Johannes~H Uhl, and Craig~A
  Knoblock.
\newblock Training deep learning models for geographic feature recognition from
  historical maps.
\newblock {\em Using Historical Maps in Scientific Studies: Applications,
  Challenges, and Best Practices}, pages 65--98, 2020.

\bibitem{christophe2022neural}
Sidonie Christophe, Samuel Mermet, Morgan Laurent, and Guillaume Touya.
\newblock Neural map style transfer exploration with gans.
\newblock {\em International Journal of Cartography}, 8(1):18--36, 2022.

\bibitem{courtial2023deriving}
Azelle Courtial, Guillaume Touya, and Xiang Zhang.
\newblock Deriving map images of generalised mountain roads with generative
  adversarial networks.
\newblock {\em International Journal of Geographical Information Science},
  37(3):499--528, 2023.

\bibitem{crampton1995ethics}
Jeremy Crampton.
\newblock The ethics of gis.
\newblock {\em Cartography and geographic information systems}, 22(1):84--89,
  1995.

\bibitem{evans2017livemaps}
Michael~R Evans, Ahmad Mahmoody, Dragomir Yankov, Florin Teodorescu, Wei Wu,
  and Pavel Berkhin.
\newblock Livemaps: Learning geo-intent from images of maps on a large scale.
\newblock In {\em Proceedings of the 25th ACM SIGSPATIAL international
  conference on advances in geographic information systems}, pages 1--9, 2017.

\bibitem{feng2019learning}
Yu~Feng, Frank Thiemann, and Monika Sester.
\newblock Learning cartographic building generalization with deep convolutional
  neural networks.
\newblock {\em ISPRS International Journal of Geo-Information}, 8(6):258, 2019.

\bibitem{griffin2020trustworthy}
Amy~L Griffin.
\newblock Trustworthy maps.
\newblock {\em Journal of Spatial Information Science}, 2020(20):5--19, 2020.

\bibitem{griffincall}
Amy~L Griffin and RMIT Anthony~C Robinson.
\newblock Call for presentations: Icc 2023 pre-conference workshop on
  cartography and ai (mapai).
\newblock 2023.

\bibitem{he2016deep}
Kaiming He, Xiangyu Zhang, Shaoqing Ren, and Jian Sun.
\newblock Deep residual learning for image recognition.
\newblock In {\em Proceedings of the IEEE conference on computer vision and
  pattern recognition}, pages 770--778, 2016.

\bibitem{hu2022enriching}
Yingjie Hu, Zhipeng Gui, Jimin Wang, and Muxian Li.
\newblock Enriching the metadata of map images: a deep learning approach with
  gis-based data augmentation.
\newblock {\em International Journal of Geographical Information Science},
  36(4):799--821, 2022.

\bibitem{kang2022review}
Yuhao Kang, Song Gao, and Robert Roth.
\newblock A review and synthesis of recent geoai research for cartography:
  Methods, applications, and ethics.
\newblock In {\em AutoCarto 2022}, 2022.

\bibitem{kang2019transferring}
Yuhao Kang, Song Gao, and Robert~E Roth.
\newblock Transferring multiscale map styles using generative adversarial
  networks.
\newblock {\em International Journal of Cartography}, 5(2-3):115--141, 2019.

\bibitem{king2023conversation}
Michael~R King and ChatGPT.
\newblock A conversation on artificial intelligence, chatbots, and plagiarism
  in higher education.
\newblock {\em Cellular and Molecular Bioengineering}, 16(1):1--2, 2023.

\bibitem{liebrenz2023generating}
Michael Liebrenz, Roman Schleifer, Anna Buadze, Dinesh Bhugra, and Alexander
  Smith.
\newblock Generating scholarly content with chatgpt: ethical challenges for
  medical publishing.
\newblock {\em The Lancet Digital Health}, 5(3):e105--e106, 2023.

\bibitem{mai2023opportunities}
Gengchen Mai, Weiming Huang, Jin Sun, Suhang Song, Deepak Mishra, Ninghao Liu,
  Song Gao, Tianming Liu, Gao Cong, Yingjie Hu, Chris Cundy, Ziyuan Li, Rui
  Zhu, and Ni~Lao.
\newblock On the opportunities and challenges of foundation models for
  geospatial artificial intelligence, 2023.
\newblock \href {http://arxiv.org/abs/2304.06798} {\path{arXiv:2304.06798}}.

\bibitem{mclean2021risks}
Scott McLean, Gemma~JM Read, Jason Thompson, Chris Baber, Neville~A Stanton,
  and Paul~M Salmon.
\newblock The risks associated with artificial general intelligence: A
  systematic review.
\newblock {\em Journal of Experimental \& Theoretical Artificial Intelligence},
  pages 1--15, 2021.

\bibitem{mhlanga2023open}
David Mhlanga.
\newblock Open ai in education, the responsible and ethical use of chatgpt
  towards lifelong learning.
\newblock {\em Education, the Responsible and Ethical Use of ChatGPT Towards
  Lifelong Learning (February 11, 2023)}, 2023.

\bibitem{monmonier2018lie}
Mark Monmonier.
\newblock {\em How to lie with maps}.
\newblock University of Chicago Press, 2018.

\bibitem{openai2023gpt4}
OpenAI.
\newblock Gpt-4 technical report, 2023.
\newblock \href {http://arxiv.org/abs/2303.08774} {\path{arXiv:2303.08774}}.

\bibitem{robinson2022visualizing}
Anthony~C Robinson and Xi~Zhu.
\newblock Visualizing viral cartography with mapreverse.
\newblock {\em GI\_Forum}, 10(1):91--97, 2022.

\bibitem{rombach2022high}
Robin Rombach, Andreas Blattmann, Dominik Lorenz, Patrick Esser, and Bj{\"o}rn
  Ommer.
\newblock High-resolution image synthesis with latent diffusion models.
\newblock In {\em Proceedings of the IEEE/CVF Conference on Computer Vision and
  Pattern Recognition}, pages 10684--10695, 2022.

\bibitem{rudolph2023chatgpt}
J{\"u}rgen Rudolph, Samson Tan, and Shannon Tan.
\newblock Chatgpt: Bullshit spewer or the end of traditional assessments in
  higher education?
\newblock {\em Journal of Applied Learning and Teaching}, 6(1), 2023.

\bibitem{shen2022raster}
Yilang Shen, Tinghua Ai, and Rong Zhao.
\newblock Raster-based method for building selection in the multi-scale
  representation of two-dimensional maps.
\newblock {\em Geocarto International}, 37(22):6494--6518, 2022.

\bibitem{tlili2023if}
Ahmed Tlili, Boulus Shehata, Michael~Agyemang Adarkwah, Aras Bozkurt, Daniel~T
  Hickey, Ronghuai Huang, and Brighter Agyemang.
\newblock What if the devil is my guardian angel: Chatgpt as a case study of
  using chatbots in education.
\newblock {\em Smart Learning Environments}, 10(1):15, 2023.

\bibitem{uhl2021combining}
Johannes~H Uhl, Stefan Leyk, Zekun Li, Weiwei Duan, Basel Shbita, Yao-Yi
  Chiang, and Craig~A Knoblock.
\newblock Combining remote-sensing-derived data and historical maps for
  long-term back-casting of urban extents.
\newblock {\em Remote sensing}, 13(18):3672, 2021.

\bibitem{van2023chatgpt}
Eva~AM van Dis, Johan Bollen, Willem Zuidema, Robert van Rooij, and Claudi~L
  Bockting.
\newblock Chatgpt: five priorities for research.
\newblock {\em Nature}, 614(7947):224--226, 2023.

\bibitem{wilson2021five}
John~P Wilson, Kevin Butler, Song Gao, Yingjie Hu, Wenwen Li, and Dawn~J
  Wright.
\newblock A five-star guide for achieving replicability and reproducibility
  when working with gis software and algorithms.
\newblock {\em Annals of the American Association of Geographers},
  111(5):1311--1317, 2021.

\bibitem{yan2019graph}
Xiongfeng Yan, Tinghua Ai, Min Yang, and Hongmei Yin.
\newblock A graph convolutional neural network for classification of building
  patterns using spatial vector data.
\newblock {\em ISPRS journal of photogrammetry and remote sensing},
  150:259--273, 2019.

\bibitem{zarifhonarvar2023economics}
Ali Zarifhonarvar.
\newblock Economics of chatgpt: A labor market view on the occupational impact
  of artificial intelligence.
\newblock {\em Available at SSRN 4350925}, 2023.

\bibitem{zhao2021deep}
Bo~Zhao, Shaozeng Zhang, Chunxue Xu, Yifan Sun, and Chengbin Deng.
\newblock Deep fake geography? when geospatial data encounter artificial
  intelligence.
\newblock {\em Cartography and Geographic Information Science}, 48(4):338--352,
  2021.

\bibitem{zhuo2023exploring}
Terry~Yue Zhuo, Yujin Huang, Chunyang Chen, and Zhenchang Xing.
\newblock Exploring ai ethics of chatgpt: A diagnostic analysis.
\newblock {\em arXiv preprint arXiv:2301.12867}, 2023.

\end{thebibliography}

\vskip 1in
\section*{Appendix}
\begin{table}[h]
\label{tab:appendix}
\centering
\caption{\textit{A list of \textit{Place} and \textit{Description} that we used in prompts for generating maps.}}
\begin{tabular}{@{}ll@{}}
\toprule
Place                   & Description                   \\ \midrule
on the table            & with black and white      \\
on the floor            & with pink and purple      \\
on the countertop       & with gold and silver      \\
on the desk             & with red and orange       \\
on the mat              & with blue and green       \\
on the tile             & with yellow and orange    \\
on the carpet           & with purple and green     \\
on the pavement         & with brown and beige      \\
on the sidewalk         & with gray and silver      \\
on the driveway         & with turquoise and blue   \\
on the parking lot      & with maroon and navy      \\
on the runway           & with pastel colors        \\
on the tarmac           & with neon colors          \\
on the tennis court     & with earth tones          \\
on the basketball court & with autumn colors        \\
on the soccer field     & with spring colors        \\
on the football field   & with summer colors        \\
on the baseball diamond & with winter colors        \\
on the hockey rink      & with metallic shades      \\
on the golf course      & with bright colors        \\
on the beach sand       & with cool colors          \\
on the pool deck        & with warm colors          \\
on the dance floor      & with muted colors         \\
on the stage            & with contrasting colors   \\
on the screen           & with complementary colors \\
on the canvas           & with analogous colors     \\
on the whiteboard       & with monochromatic colors \\
on the chalkboard       & with dark shades          \\
on the notepad          & with light shades         \\
on the easel            & with vibrant hues         \\ \bottomrule
\end{tabular}
\end{table}
\end{document}